\documentclass{article} 
\usepackage{amsmath}
\usepackage{graphicx}

\begin{document}

\begin{center}

\LARGE{Energy Gaps in a Spacetime Crystal}

\vspace{5mm}

\small{L.P. Horwitz$^{a,b,c}$ and E.Z. Engelberg$^c$}

\vspace{5mm}

\begin{footnotesize}
$^a$ School of Physics, Tel Aviv University, Ramat Aviv 69978, Israel \\
$^b$ Department of Physics, Ariel University Center of Samaria, Ariel 40700, Israel \\
$^c$ Department of Physics, Bar Ilan University, Ramat Gan 52900, Israel 
\end{footnotesize}

\vspace{5mm}

\end{center}

\begin{abstract}
This paper presents an analysis of the band structure of a spacetime potential lattice created by a standing electromagnetic wave. We show that there are energy band gaps. We estimate the effect, and propose a measurement that could confirm the existence of such phenomena.
\end{abstract}

\vspace{5mm}

\begin{small}
PACS:
03.30.+p  (special relativity),
71.20.Mq, 71.20.Nr (band structure),
71.10.-w (theory and models),
71.20.-b (crystalline solids),
85.40.-e (microelectronics)
\end{small}

\vspace{5mm}

\section{\normalsize{Introduction}}

Based upon work done by Fock\cite{fo37}, Stueckelberg\cite{st41}\cite{st42}, and Feynman\cite{fe50} in the first half of the previous century, Horwitz and Piron\cite{ho73}, and Fanchi and Collins\cite{fa78}, constructed a quantum relativistic theory in which Einstein's covariant time is considered as a dynamical variable. The evolution of a system is then parametrized by a universal invariant $\tau$ identified with Newton's time.

In this theory, the covariant wave function $\psi_\tau\left(\vec{x},t\right)$, which evolves according to the Stueckelberg-Schrodinger equation, is coherent in time as well as space variables. It provides a simple and straightforward description of interference in time\cite{ho76}, in agreement with the recent experiment of Lindner et al\cite{li05}. 

The objective of this paper is to present an analysis of the band structure of a spacetime potential lattice, created by a standing electromagnetic wave, using the Hamiltonian originally proposed by Stueckelberg (this Hamiltonian has been shown to lead to the covariant Lorentz force\cite{ho84}), and first-order perturbation theory.

\vspace{5mm}

\section{\normalsize{Mass Shift in Spacetime Lattice}}

We begin from the electromagnetic, proper-time-independent, Stueckelberg Hamiltonian 

\begin{equation}
H=\frac{\left(p_\mu-eA_\mu\right)\left(p^\mu-eA^\mu\right)}{2M}
\end{equation}

which can be written as

\begin{equation}
H = \frac{p_\mu p^\mu}{2M} - \frac{ie\hbar}{M}A^\mu \frac{\partial}{\partial x^\mu} + \frac{e^2}{2M}A_\mu A^\mu
\end{equation}

The first term is the unperturbed Hamiltonian (with spectrum corresponding to the mass of the particle). The second is a perturbation of the first order in $A^\mu$, and the third term is a perturbation of the second order in $A^\mu$.

We start by finding the first order perturbation of the first order term in $A^\mu$. Following methods used in nonrelativistic solid-state physics (see, for example, \cite{ra61}), our unperturbed wave functions are

\begin{equation}
\psi_k = \frac{1}{\sqrt{v}}e^{i k_\sigma x^\sigma}
\end{equation}

The mass shifts due to the perturbation are found by calculating the eigenvalues of the perturbation operator

\begin{equation}
\left|V - \epsilon I\right| = \left|
\begin{array}{ccc}
V_{k_1,k_1} - \epsilon & V_{k_1,k_2} & \ldots \\
V_{k_2,k_1} & V_{k_2,k_2} - \epsilon & \ldots \\
\vdots & \vdots & \ddots
\end{array}
\right| = 0
\end{equation}

with 

\begin{equation}
V_{k,k'} = \int \psi_k^* V \psi_{k'} d^4x
\label{perturbation_operator}
\end{equation}

For the first order term in $A^\mu$, this gives us

\begin{equation}
V_{k,k'} = \int\psi_k^*\frac{ie\hbar}{M}A^\mu\frac{\partial}{\partial x^\mu}\psi_{k'} d^4x
\end{equation}

We assume that the electromagnetic wave has the form

\begin{equation}
A^\mu=\left(
\begin{array}{c}
0 \\ A\sin\left(\omega_\gamma t\right)\cos\left(k_\gamma z\right) \\ 0 \\ 0
\end{array}
\right)
\label{electromagnetic_wave}
\end{equation}

Then,

\begin{align}
V_{k,k'} &= \frac{iek_xA\hbar}{4vM} \int e^{-iK_\sigma x^\sigma} \cdot \\ \nonumber
& \cdot \left(e^{i\left(\omega_\gamma t+k_\gamma z\right)} + e^{i\left(\omega_\gamma t-k_\gamma z\right)} - e^{i\left(-\omega_\gamma t+k_\gamma z\right)} - e^{i\left(-\omega_\gamma t-k_\gamma z\right)}\right) d^4x
\end{align}

with

\begin{equation}
K_\sigma = k_\sigma ' - k_\sigma
\end{equation}

This can be nonzero only when

\begin{equation}
K^\sigma =
\left(
\begin{array}{c}
\frac{\omega_\gamma}{c} \\ 0 \\ 0 \\ k_\gamma
\end{array}
\right)
or
\left(
\begin{array}{c}
\frac{\omega_\gamma}{c} \\ 0 \\ 0 \\ -k_\gamma
\end{array}
\right)
or
\left(
\begin{array}{c}
\frac{-\omega_\gamma}{c} \\ 0 \\ 0 \\ k_\gamma
\end{array}
\right)
or
\left(
\begin{array}{c}
\frac{-\omega_\gamma}{c} \\ 0 \\ 0 \\ -k_\gamma
\end{array}
\right)
\end{equation}

We define the edge of a Brillouin zone as the collection of groups of degenerate states (identical $m^2c^4 = E^2 - c^2p_z^2$), in which the distance between the points, in the $\left(E,cp_z\right)$ space, can be written as $\hbar ck_\gamma \left(n_E\widehat{E}+n_p\widehat{cp_z}\right)$, with integer $n_E$ and $n_p$.

The rank of the edge of the Brillouin zone is determined by filling in the lowest possible absolute values of integer $n_E$ and $n_p$. Thus, the edges of the first five Brillouin zones are given by the following $\left(n_E,n_p\right)$:

\begin{equation}
\begin{array}{c}
\left(n_E=\pm 1, n_p=0\right) or \left(n_E=0, n_p=\pm 1\right) \\
\left(n_E=\pm 1, n_p=\pm 1\right) \\
\left(n_E=\pm 2, n_p=0\right) or \left(n_E=0, n_p=\pm 2\right) \\
\left(n_E=\pm 2, n_p=\pm 1\right) or \left(n_E=\pm 1, n_p=\pm 2\right) \\
\left(n_E=\pm 2, n_p=\pm 2\right)
\end{array}
\begin{array}{c}
\text{1st Brillouin zone} \\
\text{2nd Brillouin zone} \\
\text{3rd Brillouin zone} \\
\text{4th Brillouin zone} \\
\text{5th Brillouin zone}
\end{array}
\end{equation}

We see that we can have a mass gap only along the edges of the 2nd Brillouin zone. However, the 2nd Brillouin zone lies entirely on the light cone, where it is not likely to find a massive charged particle.

\vspace{5mm}

\section{\normalsize{Mass Shift at the Edges of the 3rd and 5th Brillouin Zones}}

We now repeat the process for the first order perturbation of the second order term in $A^\mu$. Our unperturbed wave functions are as before, so

\begin{equation}
V_{k,k'} = \frac{1}{v} \int \psi_k^* \frac{e^2}{2M} A_\mu A^\mu \psi_{k'} d^4x
\end{equation}

Substituting eq. \ref{electromagnetic_wave} for $A^\mu$, this becomes

\begin{align}
V_{k,k'} &= \frac{e^2A^2}{8vM} \int e^{i K_\sigma r^\sigma} \cdot \\ \nonumber
& \left( 1 - \frac{1}{2}\left(e^{i2\omega_\gamma t}+e^{-i2\omega_\gamma t}-e^{i2k_\gamma z}-e^{-i2k_\gamma z}\right) - \right. \\ \nonumber
& \left. - \frac{1}{4} \left( e^{i\left(2\omega_\gamma t+2k_\gamma z\right)} + e^{i\left(2\omega_\gamma t-2k_\gamma z\right)} + e^{i\left(-2\omega_\gamma t+2k_\gamma z\right)} + e^{i\left(-2\omega_\gamma t-2k_\gamma z\right)} \right) \right) d^4x
\end{align}

This can be nonzero only when:

\begin{equation}
K^\sigma = 
\left(
\begin{array}{c}
0 \\ 0 \\ 0 \\ 0
\end{array}
\right)
\end{equation}

or

\begin{equation}
K^\sigma =
\left(
\begin{array}{c}
\frac{2\omega_\gamma}{c} \\ 0 \\ 0 \\ 0
\end{array}
\right)
or
\left(
\begin{array}{c}
\frac{-2\omega_\gamma}{c} \\ 0 \\ 0 \\ 0
\end{array}
\right)
or
\left(
\begin{array}{c}
0 \\ 0 \\ 0 \\ 2k_\gamma
\end{array}
\right)
or
\left(
\begin{array}{c}
0 \\ 0 \\ 0 \\ -2k_\gamma
\end{array}
\right)
\end{equation}

or

\begin{equation}
K^\sigma =
\left(
\begin{array}{c}
\frac{2\omega_\gamma}{c} \\ 0 \\ 0 \\ 2k_\gamma
\end{array}
\right)
or
\left(
\begin{array}{c}
\frac{2\omega_\gamma}{c} \\ 0 \\ 0 \\ -2k_\gamma
\end{array}
\right)
or
\left(
\begin{array}{c}
\frac{-2\omega_\gamma}{c} \\ 0 \\ 0 \\ 2k_\gamma
\end{array}
\right)
or
\left(
\begin{array}{c}
\frac{-2\omega_\gamma}{c} \\ 0 \\ 0 \\ -2k_\gamma
\end{array}
\right)
\end{equation}

\begin{figure}
	\centering
		\includegraphics[width=100mm]{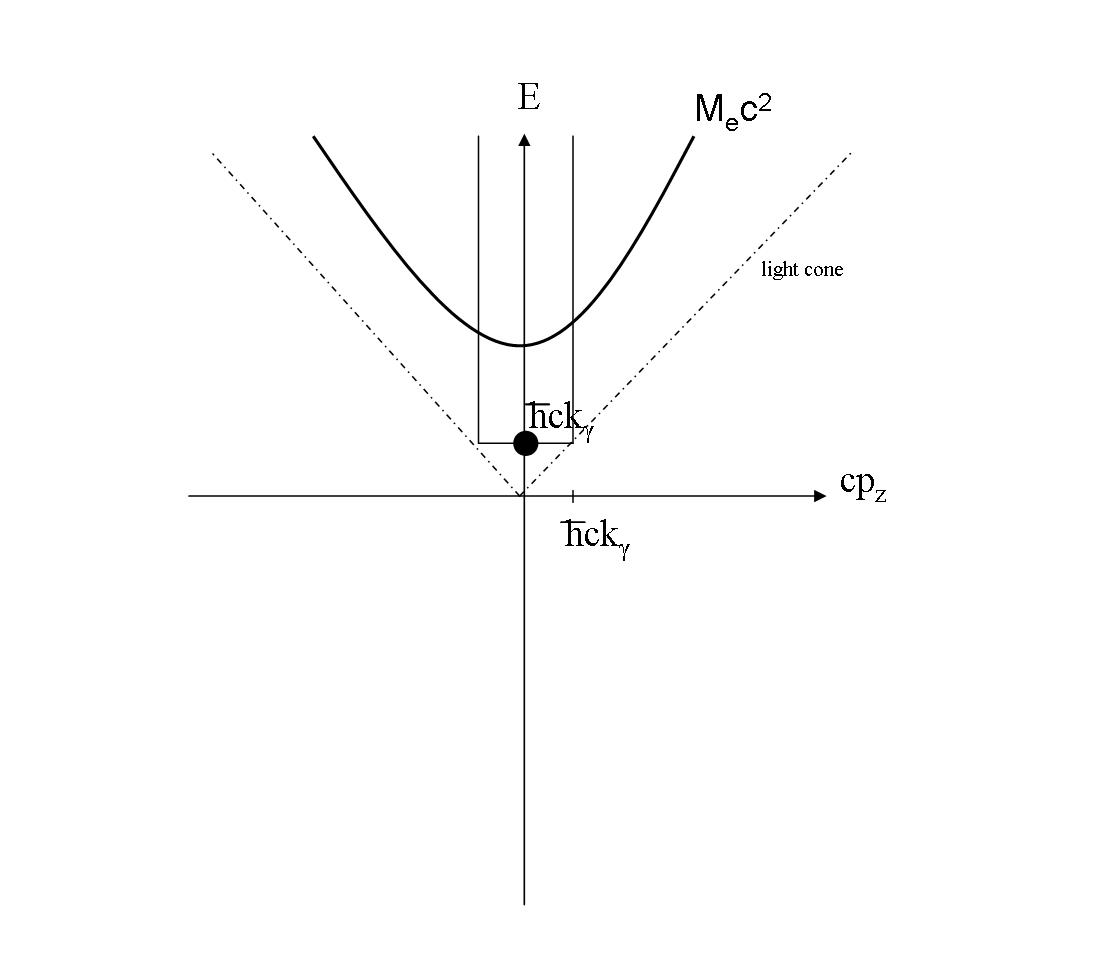}
	\caption{Edge of 3rd Brillouin Zone in Spacetime Lattice (within the light cone)}
	\label{fig:3rd_brillouin_zone}
\end{figure}

This gives us mass gaps along the edges of the 3rd and 5th Brillouin zone. The 5th Brillouin zone lies entirely on the light cone, where we are also not likely to find a massive charged particle, so our main interest lies in the gap at the edge of the 3rd zone, illustrated in figure \ref{fig:3rd_brillouin_zone}. The edges of this zone are defined by the line equations

\begin{align}
E &= \pm\hbar ck_\gamma \\
cp_z &= \pm\hbar ck_\gamma
\label{line_equation}
\end{align}

Along these lines, except for four points (where the edge of the Brillouin zone crosses the light cone), the degenerate states come in pairs, between which $K^\sigma$ is parallel to either the energy or the momentum axis. Therefore, to find the magnitude of the gap on the edge of the 3rd zone, we solve for the eigenvalues

\begin{equation}
\left|V - \epsilon_3 I\right| = \left|\quad
\begin{array}{cccc}
& \vline & k_{3\leftarrow} & k_{3\rightarrow} \\
\hline
k_{3\leftarrow} & \vline & 4\beta - \epsilon_3 & 2\beta \\
k_{3\rightarrow} & \vline & 2\beta & 4\beta - \epsilon_3 \\
\end{array}
\quad\right| = 0
\end{equation}

or

\begin{equation}
\left|V - \epsilon_3 I\right| = \left|\quad
\begin{array}{cccc}
& \vline & k_{3\downarrow} & k_{3\uparrow} \\
\hline
k_{3\downarrow} & \vline & 4\beta - \epsilon_3 & -2\beta \\
k_{3\uparrow} & \vline & -2\beta & 4\beta - \epsilon_3 \\
\end{array}
\quad\right| = 0
\end{equation}

with

\begin{equation}
\beta = \frac{e^2A^2}{32M}
\label{beta}
\end{equation}

We obtain

\begin{equation}
\epsilon_{3\pm} = 4\beta \pm 2\beta \\
\label{mass_shift}
\end{equation}

\vspace{5mm}

\section{\normalsize{Energy Gaps}}

\begin{figure}
	\centering
		\includegraphics[width=100mm]{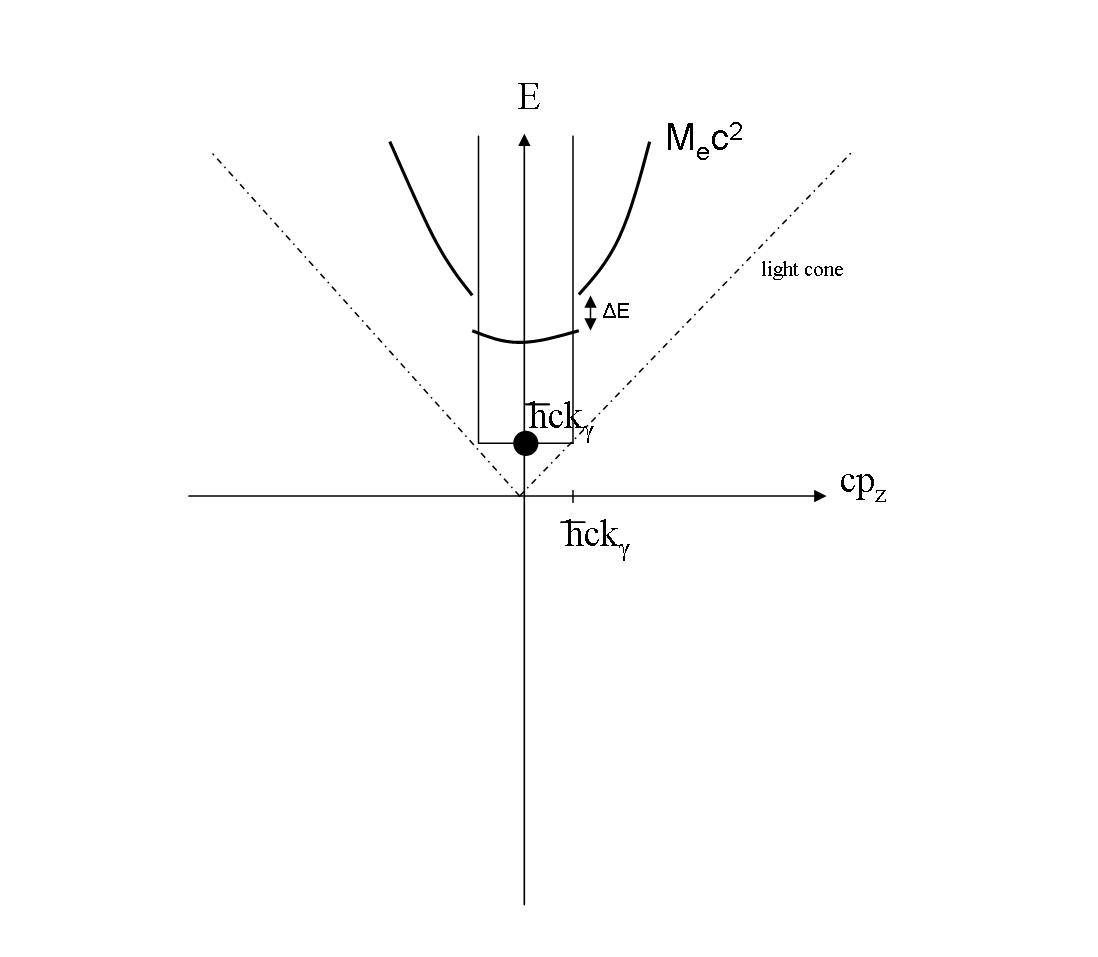}
	\caption{Mass Curve Splitting and Energy Gaps at Edge of 3rd Brillouin Zone}
	\label{fig:energy_gaps}
\end{figure}

We assume that the mass of an electron in the spacetime lattice is constrained\footnote{We assume that interaction with other fields, as well as self interaction, would provide a mechanism for the electron mass to have stability in the neighborhood of its measured value, in the absence of relatively strong perturbations. A covariant Lee-Friedrichs model was worked out as a model for a stabilizing mechanism \cite{ho95}.} to a narrow range around the mass of a free electron, $M_e$. The calculated mass gaps will be manifested as a splitting of the mass curve near the edges of the Brillouin zones, resulting in energy and/or momentum gaps, as illustrated in figure \ref{fig:energy_gaps}.

The mass equation at the edge of the 3rd Brillouin zone, after the splitting, is

\begin{equation}
\left(M_e c^2 + \epsilon_{3\pm}\right)^2 = E^2 - c^2p_z^2
\end{equation}

Substituting the positive $cp_z$ value in eq. \ref{line_equation}, we get (for positive energies)

\begin{equation}
E = \sqrt{\left(M_e c^2 + \epsilon_{3\pm}\right)^2 + \left(\hbar c k_\gamma\right)^2}
\end{equation}

If (as in the estimate below) the second term in the square root is small compared to the first one, and $\epsilon_{3\pm}$ small compared to $M_e c^2$, this becomes

\begin{equation}
E = M_e c^2 + \epsilon_{3\pm} + \frac{\left(\hbar k_\gamma\right)^2}{2M_e}
\label{energy}
\end{equation}

The vector potential amplitude $A$ is related to the intensity of the electromagnetic beam creating the lattice as

\begin{equation}
A = \frac{\sqrt{2I}}{\sqrt{c\epsilon_0}\omega_\gamma}
\label{vector_potential}
\end{equation}

Combining eqs. \ref{beta}, \ref{mass_shift}, \ref{energy}, and \ref{vector_potential}, we have

\begin{equation}
E = M_e c^2 + \left(2 \pm 1\right) \frac{e^2 I \lambda^2}{16M_e c^3 \epsilon_0} + \frac{\hbar^2}{2M_e \lambda^2}
\end{equation}

with $\lambda$ the wavelength of the beam creating the lattice.

This means that all kinetic energy values of the electron between

\begin{equation}
E_- = \frac{\hbar^2}{2M_e \lambda^2} + \frac{e^2 I \lambda^2}{16M_e c^3 \epsilon_0}
\end{equation}

and 

\begin{equation}
E_+ = \frac{\hbar^2}{2M_e \lambda^2} + 3 \frac{e^2 I \lambda^2}{16M_e c^3 \epsilon_0}
\end{equation}

are forbidden inside the lattice.

For example, if the electromagnetic beam creating the lattice has a wavelength of $589 nm$, and an intensity of $3.13\cdot 10^{12} \frac{W}{cm^2}$, we have

\begin{align}
&\frac{\hbar^2}{2M_e \lambda^2} = 1.10\cdot 10^{-7}eV \\ \nonumber
&\frac{e^2 I \lambda^2}{16M_e c^3 \epsilon_0} = 0.5eV
\end{align}

Therefore

\begin{align}
E_- &= 0.5eV \\ \nonumber
E_+ &= 1.5 eV
\end{align}

meaning that all kinetic energies between 0.5eV and 1.5eV are forbidden. 

\vspace{5mm}

\section{\normalsize{Discussion and Conclusions}}

We have shown that one can conceive of a spacetime lattice in analogy to the crystals of nonrelativistic solid state physics, associated with a standing wave in a cavity. The resulting wave functions, according to the Schr\"odinger-Stueckelberg equation, are Bloch waves with energy gaps. Our estimates indicate that this phenomenon can be detected in laboratory measurements.

\end{document}